\documentclass[prl,a4paper,nofootinbib,twocolumn,showpacs,preprintnumbers,amsmath,amssymb,floatfix,amstex,superscriptaddress]{revtex4}
\newcommand{\ket}[1]{|#1\rangle}                    %
\newcommand{\bra}[1]{\langle #1}                    %
\usepackage{graphicx,graphics,wrapfig,rotating}         % Include figure files
\usepackage{dcolumn}                            % Align table columns on decimal point
\usepackage{bbm, bm,fancybox}                       % bold math
\usepackage{times,euscript,eufrak,oldgerm}              %
\usepackage[german,english]{babel}          %
\usepackage{psfrag}
\usepackage{color}

\begin{document}
\title{Scaling laws for the decay of multiqubit entanglement} %
\author{L. Aolita}
\affiliation{%
Instituto de F\'\i sica, Universidade Federal do Rio de Janeiro. Caixa Postal
68528, 21941-972 Rio de Janeiro, RJ, Brasil\\
}
\author{R. Chaves}
\affiliation{%
Instituto de F\'\i sica, Universidade Federal do Rio de Janeiro. Caixa Postal
68528, 21941-972 Rio de Janeiro, RJ, Brasil\\
}
\author{D. Cavalcanti}
\affiliation{ICFO-Institut de Ciencies Fotoniques, Mediterranean
Technology Park, 08860 Castelldefels (Barcelona), Spain}
\author{A. Ac\'in}
\affiliation{ICFO-Institut de Ciencies Fotoniques, Mediterranean
Technology Park, 08860 Castelldefels (Barcelona), Spain}
\affiliation{ICREA-Instituci\'o Catalana de Recerca i Estudis
Avan\c cats, Lluis Companys 23, 08010 Barcelona, Spain}

\author{L. Davidovich}
\affiliation{%
Instituto de F\'\i sica, Universidade Federal do Rio de Janeiro. Caixa Postal
68528, 21941-972 Rio de Janeiro, RJ, Brasil\\
}
\date{\today}%
\begin{abstract}
We investigate the decay of entanglement of generalized $N$-particle
Greenberger-Horne-Zeilinger (GHZ) states
interacting with independent reservoirs. Scaling laws for the decay
of entanglement and for its finite-time extinction (sudden death)
are derived for different types of reservoirs. The latter is found to
increase with the number of particles. However, entanglement becomes
arbitrarily small, and therefore useless as a resource, much before
it completely disappears, around a time which is inversely
proportional to the number of particles.  We also show that the
decay of multi-particle GHZ states can generate bound entangled
states. %%
\end{abstract}
\pacs{03.67.-a, 03.67.Mn, 03.65.Yz}
\maketitle
%%%%%%%%%%%%%%%%%%%%%%%%%%%%%%%%%%%%%%%%%%%%%%%%%%%%%%%%%%%%%%%%%%%%%%%%%%%%%%%%%%%%%%%%%%%%%%%%%%%%%%%%%%%%%%%%%%%%%%%%%%%%%%%%
{\bf Introduction.} Entanglement has been identified as a key
resource for many potential practical applications, such as
quantum computation, quantum teleportation and quantum
cryptography \cite{chuang00}. Being it a resource, it is of
fundamental importance to study the entanglement properties of
quantum states in realistic situations, where the system
unavoidably loses its coherence due to interactions with the
environment. In this context a peculiar dynamical feature of
entangled states has been experimentally confirmed for the case of
two qubits (two-level systems) \cite{Science}: even when the
constituent parts of an entangled state decay asymptotically in
time, entanglement may disappear at a finite time
\cite{Simon&Kempe,
Andre,Duer&Briegel,sudden-death,Yu&Eberly,Al-Qasimi&James,Yu&Eberly07}.
The phenomenon of finite-time disentanglement, also known as
entanglement sudden death (ESD)
\cite{Science,Yu&Eberly,Al-Qasimi&James,Yu&Eberly07}, illustrates
the fact that the global behavior of an entangled system, under
the effect of local environments, may be markedly different from
the individual and local behavior of its constituents.

\par Since the speed-up gained when using quantum-mechanical systems, instead of classical ones, to process information is
only considerable in the limit of large-scale information
processing, it is fundamental to understand the scaling properties
of disentanglement for multiparticle systems. Important steps in
this direction were given in Refs.
\cite{Simon&Kempe,Andre,Duer&Briegel}. In particular, it was shown
in Ref. \cite{Simon&Kempe} that  balanced Greenberger-Horne-Zeilinger
(GHZ) states, $\ket{\Psi}=(\ket{0}^{\otimes N}+\ket{1}^{\otimes
N})/\sqrt{2}$, subject to the action of individual depolarization
\cite{chuang00}, undergo ESD, that the last bipartitions to loose
entanglement are the most balanced ones, and that the time at which
such entanglement disappears grows with the number $N$ of particles
in the system. Soon afterwards it was shown in Ref.~\cite{Duer&Briegel} that
the first bipartitions to loose entanglement are the least balanced
ones (one particle vs. the others), the time at which this happens
decreasing with $N$.
A natural question arises from these considerations: {\it is the ESD
time a truly physically-relevant quantity to assess the robustness
of multi-particle entanglement?}

\par In this paper we show that, for an important family of genuine-multipartite entangled states, {\it the answer is no}.
For several kinds of decoherence, we derive analytical expressions
for the time of disappearance of bipartite entanglement, which is
found to increase with $N$. However, we show that the time at
which bipartite entanglement becomes arbitrarily small decreases
with the number of particles, independently of ESD. This implies
that for multi-particle systems, the amount of
entanglement can become too small for any practical application
long before it vanishes.  In addition, for some specific cases, we
characterize not only the sudden-death time of bipartite
entanglement but we can also attest full separability of the states
in question. As a byproduct we show that in several cases the
action of the environment can naturally lead to bound entangled
states \cite{Hor}, in the sense that, for a period of time, it is
not possible to extract pure-state entanglement from the system
through local operations and classical communication, even though the state is still entangled.

The exemplary states we take to analyze the robustness of
multipartite entanglement are generalized GHZ states:
\begin{equation}
\label{state}
\ket{\Psi_0}\equiv\alpha\ket{0}^{\otimes N}+\beta\ket{1}^{\otimes N},
\end{equation}
with $\alpha$ and $\beta\in\mathbb{C}$ such that
$|\alpha|^2+|\beta|^2=1$. Therefore, our results also constitute a
generalization of those of Refs. \cite{Simon&Kempe,Duer&Briegel}.
Although \eqref{state} represents just a restricted class of states,
the study of its entanglement properties is important in its own
right: these can be seen as simple models of the Schr\"odinger-cat
state \cite{sch}, they are crucial for communication problems
\cite{GHZuse}, and such states have been experimentally produced in
atomic and photonic systems of up to $N=6$ \cite{GHZexp}.

%and used in photonic \cite{GHZexp1} and atomic \cite{GHZexp2}
%systems of up to $N=6$.

%%%%%%%%%%%%%%%%%%%%%%%%%%%%%%%%%%%%%%%%%%%%%%%%%%%%%%%%%%%%%%%%%%%%%%%%%%%%%%%%%%%%%%%%%%%%%%%%%%%%
\par{\bf Decoherence models.}  We consider three paradigmatic types of noisy channels: depolarization,
dephasing, and a thermal bath at arbitrary temperature
(generalized amplitude-damping channel). We consider $N$ qubits of ground state $\ket{0}$ and excited state
$\ket{1}$ without mutual interaction, each one individually
coupled  to its own noisy environment. The dynamics of the i-th
qubit, $1\le i\le N$, is governed by a master equation that gives
rise to a completely positive  trace-preserving map (or channel)
${\cal E}_i$ describing the evolution as $\rho_i={\cal
E}_i\rho_{0_{i}}$, where $\rho_{0_{i}}$ and $\rho_i$ are,
respectively, the initial and evolved reduced states of the i-th
subsystem.

\par The generalized amplitude-damping channel (GAD) is given, in the Born-Markov approximation, via its Kraus representation
as \cite{chuang00,Yu&Eberly07}
\begin{equation}
\label{GADC}
{\cal E}^{GAD}_i\rho_{i}=E_0\rho_{i} E_{0}^{\dagger}+E_1\rho_{i} E_{1}^{\dagger}+E_2\rho_{i} E_{2}^{\dagger}+E_3\rho_{i} E_{3}^{\dagger};
\end{equation}
with
$E_0\equiv\sqrt{\frac{\overline{n}+1}{2\overline{n}+1}}(\ket{0}\bra{0}|+\sqrt{1-p}\ket{1}\bra{1}|)$,
$E_1\equiv\sqrt{\frac{\overline{n}+1}{2\overline{n}+1}p}\ket{0}\bra{1}|$,
$E_2\equiv\sqrt{\frac{\overline{n}}{2\overline{n}+1}}(\sqrt{1-p}\ket{0}\bra{0}|+\ket{1}\bra{1}|)$
and
$E_3\equiv\sqrt{\frac{\overline{n}}{2\overline{n}+1}p}\ket{1}\bra{0}|$
being its Kraus operators. Here $\overline{n}$ is the mean number
of excitations in the bath, $p\equiv p(t)\equiv
1-e^{-\frac{1}{2}\gamma(2\overline{n}+1)t}$ is the probability of
the qubit exchanging a quantum with the bath at  time $t$, and
$\gamma$ is the zero-temperature dissipation rate.
Channel~(\ref{GADC}) is a generalization to finite temperature of
the purely dissipative amplitude damping channel (AD), which is
obtainen from (\ref{GADC}) in the zero- temperature limit
$\overline{n}=0$. On the other hand, the purely diffusive case is
obtained from (\ref{GADC}) in the composite limit
$\overline{n}\rightarrow\infty$, $\gamma\rightarrow0$, and
$\overline{n}\gamma=\Gamma$, where $\Gamma$ is the diffusion
constant.

\par The depolarizing channel (D) describes the situation in which the i-th qubit remains untouched with probability
$1-p$, or is depolarized - meaning that its
state  is taken to the maximally mixed state (white noise) - with probability $p$. It can
be expressed as
\begin{equation}
\label{DC} {\cal E}^{D}_i\rho_i=(1-p)\rho_i+(p){\bm 1}/2,
\end{equation}
where ${\bm 1}$ is the identity operator.
\par Finally, the phase damping (or dephasing) channel (PD) represents the situation in which there is loss of quantum information
 with probability $p$, but without any energy exchange. It is defined as
\begin{equation}
\label{PD}
{\cal E}^{PD}_i\rho_i=(1-p)\rho_i+p\big(\ket{0}\bra{0}|\rho_i\ket{0}\bra{0}|+\ket{1}\bra{1}|\rho_i\ket{1}\bra{1}|\big).
\end{equation}
The parameter $p$ in channels (\ref{GADC}), (\ref{DC}) and (\ref{PD}) is a convenient parametrization of time: $p=0$ refers
to the initial time 0 and $p=1$ refers to the asymptotic $t\rightarrow\infty$ limit.

\par The density matrix corresponding to state (\ref{state}),
$\rho_0\equiv\ket{\Psi_0}\bra{\Psi_0}|\equiv|\alpha|^2(\ket{0}\bra{0}|)^{\otimes
N}+|\beta|^2(\ket{1}\bra{1}|)^{\otimes N}
+\alpha\beta^*(\ket{0}\bra{1}|)^{\otimes
N}+\alpha^*\beta(\ket{1}\bra{0}|)^{\otimes N}$,
 evolves in time into a mixed state $\rho$ given simply by the composition of all $N$ individual maps: $
\rho\equiv {\cal E}_1{\cal E}_2\  ...\   {\cal E}_N\rho_0$, where, in what follows, ${\cal E}_i$ will either be given by Eqs. (\ref{GADC}),
(\ref{DC}) or  (\ref{PD}).
%%%%%%%%%%%%%%%%%%%%%%%%%%%%%%%%%%%%%%%%%%%%%%%%%%%%%%%%%%%%%%%%%%%%%%%%%%%%%%%%%%%%%%%%%%%%%%%%%%%%
\par{\bf Entanglement sudden death}.
In order to pick up the entanglement features of the studied states
we will use the negativity as a quantifier of entanglement
\cite{VidWer}, defined as the absolute value of the sum of the
negative eigenvalues of the partially transposed density matrix. In general,
the negativity fails to quantify entanglement of some entangled
states (those ones with positive partial transposition) in
dimensions higher than six \cite{Peres-Horodecki}. However, for the
states considered here, their partial transposes have at most one
negative eigenvalue, and the task of calculating the negativity
reduces to a four-dimensional problem. So, in the considered cases,
the negativity brings all the relevant information about the
separability in bipartitions of the states, i.e., null negativity
means separability in the corresponding partition.

Application of channel (\ref{GADC}) to every qubit multiplies the
off-diagonal elements of $\rho_0$ by the factor $(1-p)^{N/2}$,
whereas application of channels (\ref{DC}) or (\ref{PD}), by the
factor  $(1-p)^{N}$. The diagonal terms $(\ket{0}\bra{0}|)^{\otimes
N}$ and $(\ket{1}\bra{1}|)^{\otimes N}$ in turn give rise to new
diagonal terms of the form $(\ket{0}\bra{0}|)^{\otimes
N-k}\otimes(\ket{1}\bra{1}|)^{\otimes k}$, for $1\le k < N$, and all
permutations thereof, with coefficients $\lambda_k$ given below. In
what follows we present the main results concerning the entanglement
behavior of these states.

\par \emph{Generalized amplitude-damping channel:}  Consider a bipartition $k:N-k$ of the quantum state.
For channel~(\ref{GADC}), the coefficients $\lambda_k^{GAD}$ are
given by
$\lambda_k^{GAD}\equiv|\alpha|^2x^{N-k}y^{k}+|\beta|^2w^{N-k}z^{k}$,
with $0\le  x\equiv\frac{-p\overline{n}}{2\overline{n}+1}+1,\
y\equiv\frac{p\overline{n}}{2\overline{n}+1}, \
w\equiv\frac{p(\overline{n}+1)}{2\overline{n}+1}$ and
$z\equiv\frac{-p(\overline{n}+1)}{2\overline{n}+1}+1\le 1$. From
them, the minimal eigenvalue of the states' partial transposition,
% - the negativity -
$\Lambda_k^{GAD}(p)$, is immediately obtained for the generalized
amplitude damping channel \cite{note}:
\begin{equation}
\label{NegGADC}
\Lambda_k^{GAD}(p)\equiv\delta_k-\sqrt{\delta_k^2-\Delta_k}\,.
\end{equation}
Here $\delta_k=1/2[\lambda_k^{GAD}(p)+\lambda_{N-k}^{GAD}(p)]$ and
$\Delta_k=\lambda_k^{GAD}(p)\lambda_{N-k}^{GAD}(p)-|\alpha\beta|^2(1-p)^{N}$.
%is the determinant of the partial transpose.
From~(\ref{NegGADC}) one can see that
$|\Lambda_1^{GAD}(p)|\le|\Lambda_2^{GAD}(p)|\le \ ...\
\le|\Lambda_{\frac{N}{2}}^{GAD}(p)|$, for $N$ even, and
$|\Lambda_1^{GAD}(p)|\le|\Lambda_2^{GAD}(p)|\le \ ...\
\le|\Lambda_{\frac{N-1}{2}}^{GAD}(p)|$, for $N$ odd.

\par The condition for disappearance of bipartite entanglement, $\Lambda_k^{GAD}(p)=0$,
is a polynomial equation of degree $2N$. In the purely dissipative case
$\overline{n}=0$, a simple analytical solution yields the
corresponding critical probability for the amplitude-decay channel,
$p_c^{AD}$ (with $\beta\neq 0$):
\begin{equation}
\label{ESD@T=0}
p_{c}^{AD}(k)=\min\{1,|\alpha/\beta|^{2/N}\}.
\end{equation}
For $|\alpha|<|\beta|$ probability~(\ref{ESD@T=0}) is always
smaller than 1, meaning that bipartite entanglement disappears
before the steady state is asymptotically reached. Thus,
contition~(\ref{ESD@T=0}) is the direct generalization to the
multiqubit case of the ESD condition of Refs. \cite{Yu&Eberly,Science}
for two qubits subject to amplitude damping. A remarkable feature
about contition~(\ref{ESD@T=0}) is that it displays no dependence
on the number of qubits  $k$ of the sub-partition. That is,
the negativities corresponding to bipartitions composed of different
numbers of qubits  all vanish at the same time, even though they
follow different evolutions.
In the appendix we prove that at this point the state is not only
separable according to all of its bipartitions but it is indeed
fully separable, i.e., it can be written as a convex combination
of product states.

\par For arbitrary temperature, it is enough to consider the case $k=N/2$, as the entanglement corresponding to the most balanced bipartitions is
 the last one to disappear (we take $N$ even from now on just for simplicity). For arbitrary temperature, the condition $\Lambda_{N/2}^{GAD}(p)=0$ reduces to a
 polynomial equation of degree $N$, which for the purely diffusive case yields:
\begin{equation}
\label{ESD@T=infty}
p_{c}^{Diff}(N/2)=1+2|\alpha\beta|^{2/N}-\sqrt{1+4|\alpha\beta|^{4/N}}\,.
\end{equation}

\par \emph{Depolarizing channel:} For channel~(\ref{DC}), the coefficients $\lambda_k^{D}$ of $\rho$ are given by
$\lambda_k^{D}\equiv|\alpha|^2(1-\frac{p}{2})^{N-k}(\frac{p}{2})^{k}+|\beta|^2(1-\frac{p}{2})^{k}(\frac{p}{2})^{N-k}$.
One obtains again
$\Lambda_k^{D}(p)\equiv\delta_k-\sqrt{\delta_k^2-\Delta_k}$, with
$\delta_k=1/2[\lambda_k^{D}+\lambda_{N-k}^{D}]$ and
$\Delta_k=\lambda_k^{D}\lambda_{N-k}^{D}-|\alpha\beta|^2(1-p)^{2N}$.
Also here it is easy to show that the negativity associated to the
most balanced bipartition is always higher than the others, while
the one corresponding to the least balanced partition is the
smallest one. The critical probability for the disappearance of
entanglement in the $N/2:N/2$ partition is given by:
\begin{equation}
\label{ESDDC} p_{c}^{D}(N/2)=1-(1+4|\alpha\beta|^{2/N})^{-1/2}.
\end{equation}

\par \emph{Phase damping channel:} Finally, for the phase damping channel, whereas the off-diagonal
terms of the density matrix evolve as mentioned before, all the
diagonal ones  remain the same, with $\lambda_k^{PD}\equiv
0\equiv\lambda_{N-k}^{PD}$ for $1\le k<N$.
%, so that, by the D\"ur-Cirac-Tarrach criterion \cite{Duer&Cirac}, when all
%negativities vanish the state is fully separable.
In this case,
%the negativity
%$\Lambda_k^{PD}(p)$
%\iffalse
%corresponding to the partial
%transposition according to this bipartition
%\fi
 %
 %\begin{equation}
%\label{NegPD}
$\Lambda_k^{PD}(p)\equiv-|\alpha\beta|(1-p)^{N}$.
%\end{equation}
This expression is independent of $k$, and therefore of the
bipartition, and for any $\alpha,\beta\neq0$  it vanishes only for
$p=1$, i. e., only in the asymptotic time limit, when the state is
completely separable: \emph{generalized GHZ states of the form~(\ref{state}),  subject to individual
 dephasing, never experience ESD.}

\par{\bf The environment as a creator of bound entanglement.} Some effort has been recently done in order to understand whether
bound entangled (i.e. undistillable) states naturally arise from
natural physical processes \cite{bound}. In this context, it has
been found that different many-body models present thermal bound
entangled states \cite{bound}. Here we show, in a conceptually
different approach, that bound entanglement can also appear in
dynamical processes, namely decoherence.

For all channels here considered, the property
$|\Lambda_1(p)|\le|\Lambda_2(p)|\le \ ...\
\le|\Lambda_{\frac{N}{2}}(p)|$ holds. Therefore, when
$|\Lambda_1(p)|=0$, there may still be entanglement in the global
state for some time afterwards, as detected by other partitions.
When this happens, the state, even though entangled, is separable
according to every $1:N-1$ partition, and then no entanglement can
be distilled by (single-particle) local operations.

\begin{figure}%[t]
\begin{center}
\includegraphics[width=1\linewidth]{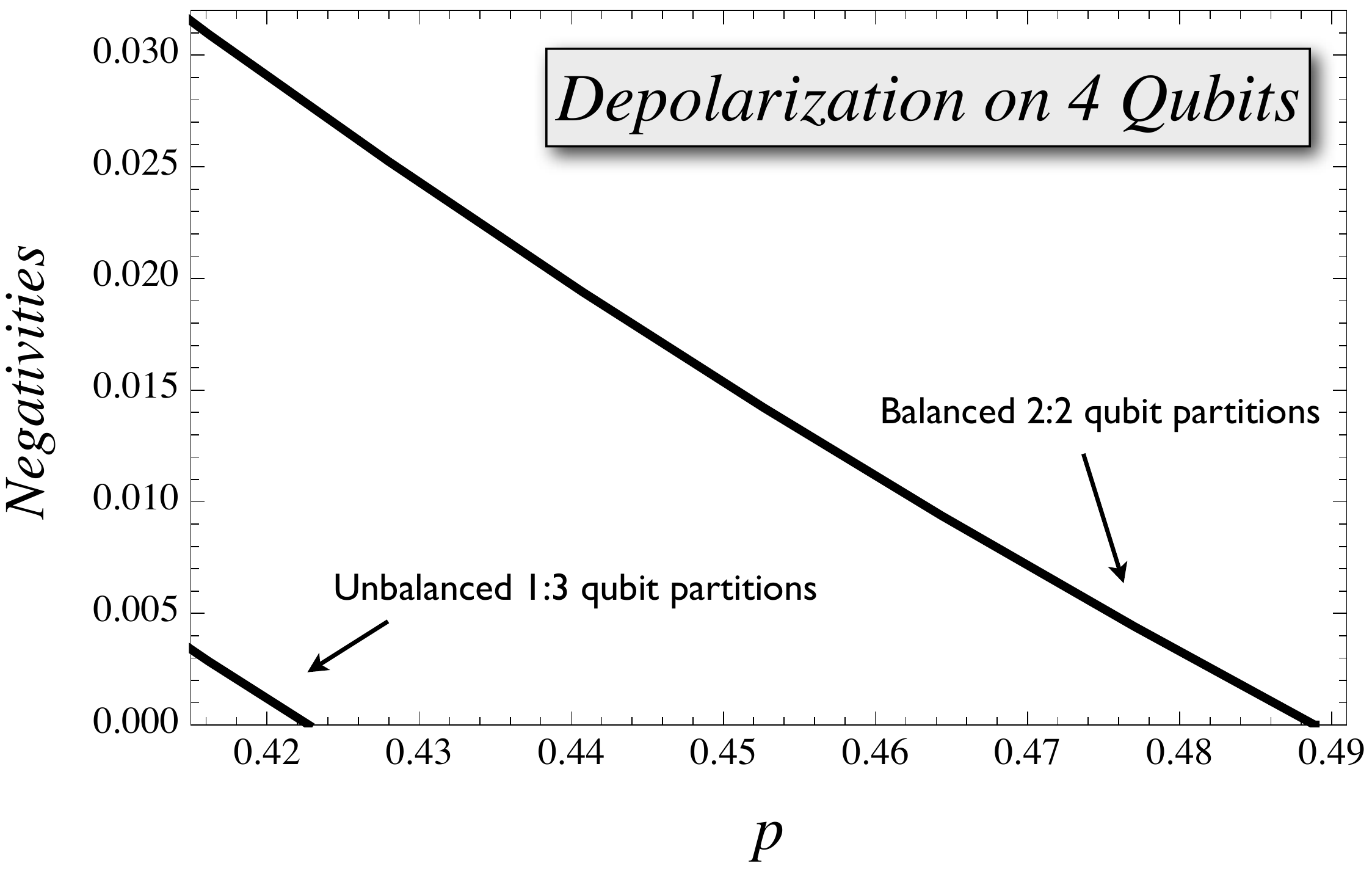}
\caption{ \label{Fig1} Negativity
as a function of $p$ for a
balanced, $\alpha=1/\sqrt{2}=\beta$, four-qubit GHZ state and independent depolarizing channels. A similar behavior is observed with channel GAD with $\overline{n}\neq0$, but the effect is not so marked (the smaller $\overline{n}$, the weaker the effect).}
\end{center}
\end{figure}

An example of this is shown in Fig.~\ref{Fig1}, where the
negativity for partitions $1:N-1$ and $N/2:N/2$ is plotted versus
$p$, for $N=4$ and $\alpha=1/\sqrt{2}=\beta$, for channel D. After
the $1:3$ negativity vanishes, the $2:2$ negativity remains
positive until $p=p_{c}^{D}(2)$ given by Eq. \eqref{ESDDC}.
Between these two values of $p$, the state is bound entangled
since it is not separable but no entanglement can be extracted
from it locally. Therefore,  the environment itself is a natural
generator of bound entanglement. Of course, this is not the case
for channels AD and PD, since for the former the state is fully
separable at $p_{c}^{AD}(k)$ (see Eq. \eqref{ESD@T=0} and
Appendix)
%when all negativities vanish together
while the latter never induces ESD.

\par{\bf Does the time of ESD really matter for large N?} Inspection of critical probabilities~(\ref{ESD@T=0}), (\ref{ESD@T=infty}) and
 (\ref{ESDDC}) shows that in all three cases $p_c$ grows with $N$. In fact, in the limit $N\rightarrow\infty$ we have, for
 $|\alpha\beta|\neq 0$, $p_{c}^{AD}(k)\rightarrow 1$, $p_{c}^{Diff}(N/2)\rightarrow 3-\sqrt{5}\approx 0.76$ and
 $p_{c}^{D}(N/2)\rightarrow 1-\frac{1}{\sqrt{5}}\approx 0.55$. \begin{figure}[b]
\begin{center}
\includegraphics[width=1\linewidth]{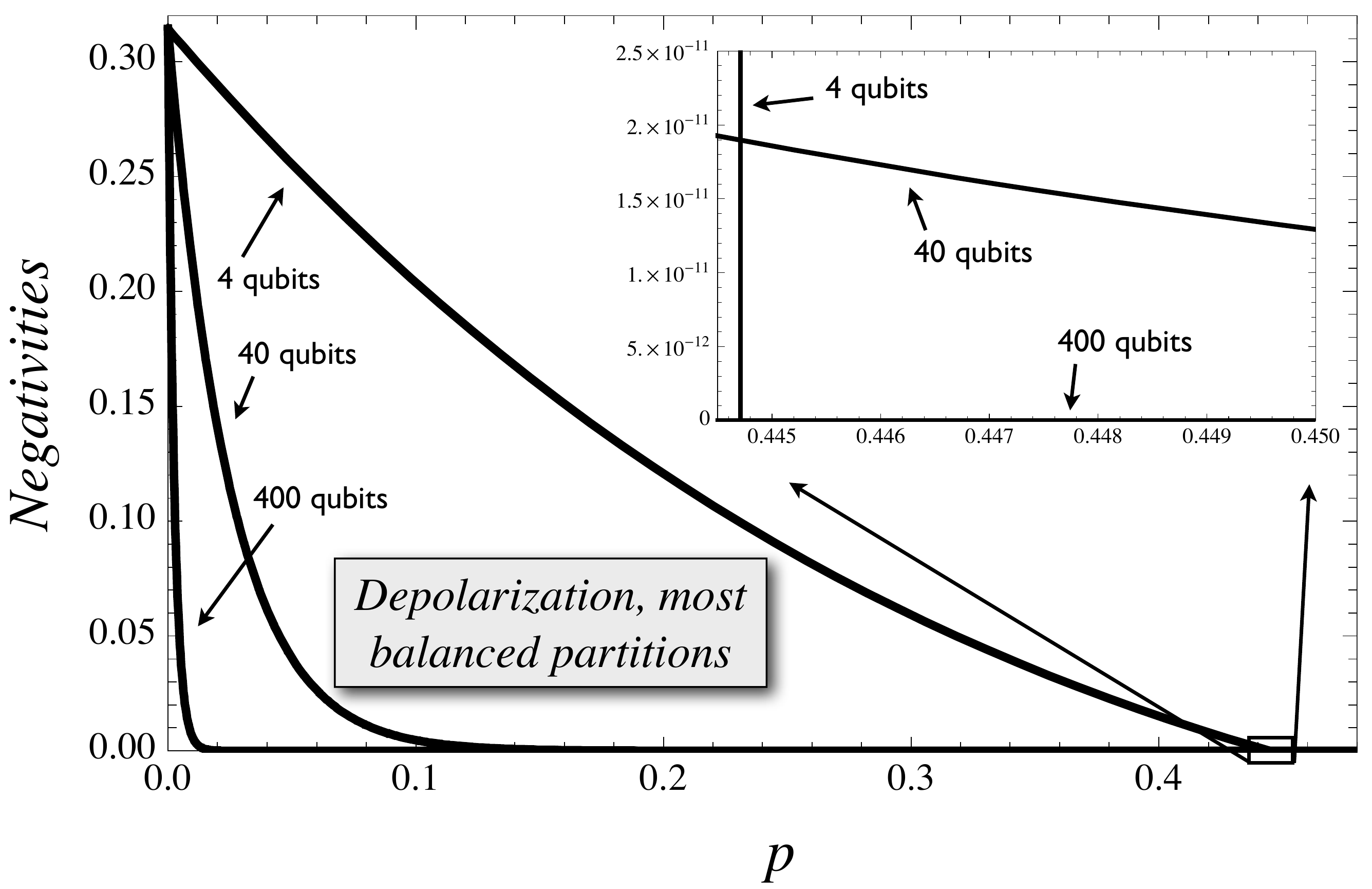}
\caption{
\label{Fig2}
Negativity versus $p$  for $N=4$, $40$ and $400$, for channel D and for the most balanced partitions. In this graphic $\alpha=1/3$ and $\beta=\sqrt{8}/3$,
but the same behavior is displayed for all other parameters and
maps. The inset shows a magnification of the region in which
$|\Lambda_{2}^{D}(p)|$ vanishes. Even though $|\Lambda_{20}^{D}(p)|$
and $|\Lambda_{200}^{D}(p)|$ cross the latter and vanish much later,
they  become orders of magnitude smaller than their initial value
long before reaching the crossing point.}
\end{center}
\end{figure}
This might be interpreted as the state's entanglement becoming more
robust when the system's size increases. However, what really
matters is not that the initial entanglement does not disappear but
that a significant fraction of it remains, either to be directly
used, or to be distilled without an excessively large overhead in
resources. The idea is clearly illustrated in Fig.~\ref{Fig2}, where
the negativity corresponding to the most balanced partitions is
plotted versus $p$ for different values of $N$. Even though the ESD time
increases with $N$, the time at which entanglement becomes
arbitrarily small decreases with it. The channel used in
Fig.~\ref{Fig2} is the depolarizing channel, nevertheless the
behavior is absolutely general, as discussed in the following.

\par For an arbitrarily small real $\epsilon>0$, and all states for which $|\alpha\beta|\neq0$,  the critical probability
$p_{\epsilon}$ at which  $\Lambda_{N/2}(p_{\epsilon})=\epsilon\Lambda_{N/2}(0)$, becomes inversely proportional to $N$
in the limit of large $N$.  For channel (\ref{GADC}), this is shown by letting $k=N/2$ in~(\ref{NegGADC}), which simplifies to
$\Lambda_{N/2}^{GAD}(p)=-|\alpha\beta|(1-p)^{N/2}+|\alpha|^2x^{N/2}y^{N/2}+|\beta|^2w^{N/2}z^{N/2}$.
For any mean bath excitation $\overline{n}$, $x^{N/2}$ and $z^{N/2}$ are at most of the same order of magnitude as  $(1-p)^{N/2}$,
whereas $y^{N/2}$ and $w^{N/2}$ are much smaller than one. Therefore, for all states such that $|\alpha\beta|\neq 0$ we
can neglect the last two terms and approximate (\ref{NegGADC}), at $k=N/2$, as $\Lambda_{N/2}^{GAD}(p)=-|\alpha\beta|^2(1-p)^{N/2}$.
We set now $\Lambda_{N/2}^{GAD}(p_{\epsilon})=\epsilon\Lambda^{GAD}_{N/2}(0)\Rightarrow\epsilon=(1-p_{\epsilon})^{N/2}\Rightarrow\log(\epsilon)=\frac{N}{2}\log(1-p_{\epsilon})$. Since $p_{\epsilon}\ll p_{c}^{GAD}(N/2)\le1$, we can approximate the logarithm on the right-hand side of the last equality by its Taylor expansion up to first order in $p_{\epsilon}$ and write $\log(\epsilon)=-\frac{N}{2}p_{\epsilon}$, implying that
%\begin{equation}
%\label{GADCinverselyprop}
$p^{GAD}_{\epsilon}\approx-(2/N)\log(\epsilon).$
%\end{equation}
Similar reasonings applied to  channels  (\ref{DC}) and (\ref{PD})
lead to
%\begin{equation}
%\label{DCinverselyprop}
$p^{D,PD}_{\epsilon}(t)\approx-(1/N)\log(\epsilon).$
%\end{equation}
These expressions
%Eqs.~(\ref{GADCinverselyprop}) and (\ref{DCinverselyprop})
assess the robustness of the state's
entanglement better than the ESD time. Much before ESD, negativity
becomes arbitrarily small. The same behavior is observed for all
studied channels, and all coefficients $\alpha$, $\beta\neq0$,
despite the fact that for some cases, like for instance for
channel (\ref{PD}), no ESD is observed. The presence of
$\log\epsilon$ in the above expression shows that our result is
quite insensitive to the actual value of $\epsilon\ll1$.
\par{\bf Conclusions}. We probed the robustness of the entanglement
of generalized GHZ states of arbitrary number of particles, $N$, subject to
 independent environments.  The states
possess in general longer ESD time, the bigger $N$, but  the time at
which such entanglement becomes arbitrarily small is inversely
proportional to $N$. The latter time  characterizes better the
robustness of the state's entanglement than the time at which ESD
itself occurs.  In several cases the action of the environment can
naturally lead to bound entangled states. An open question still
remains on how other genuinely multipartite entangled states, such
as graph states, behave. W states are expected to be more robust,
since they have always only one excitation, regardless of $N$
\cite{W}. For example, it is possible to show that, for $W$ states,
channel AD induces no ESD; however, the negativity of the
least balanced partitions decays with $1/\sqrt{N}$ \cite{Planet}.
This is another instance in which the ESD time is irrelevant to
assess the robustness of multi-particle entanglement. Our results
suggest that maintaining a significant amount of multiqubit
entanglement in macroscopic systems might be an even harder task
than believed so far.

We thank F. Mintert and A. Salles for helpful comments and FAPERJ,
CAPES, CNPQ, Brazilian Millenium Institute for Quantum
Information, EU QAP project, Spanish MEC under FIS2004-05639, and
Consolider-Ingenio QOIT projects for financial support.

\par{\bf Appendix.} Here we prove that the amplitude
damping channel leads the state \eqref{state} to a fully separable
state when all of its bipartite entanglements vanish.

The evolved state can be written as
$\rho=|\alpha|^2(\ket{0}\bra{0}|)^{\otimes N}+\rho_s$, where
$\rho_s$ is an unnormalized state. The goal is to show that
$\rho_s$ is fully separable. This is done by showing that $\rho_s$ is obtained, with a certain probability, from a fully
separable state $\sigma$ through a {\it local} positive-operator-valued measurement (POVM) \cite{chuang00}. Because only local
operations are applied, we conclude that $\rho_s$, and thus
$\rho$, must be fully separable.

The (unnormalized) state $\sigma$ is defined as
$\sigma=2^{-N}|\beta|^2\{1+|\beta/\alpha|[\frac{\alpha}{\beta}(\ket{0}\bra{1}|)^{\otimes
N}+\frac{\alpha^*}{\beta^*}(\ket{1}\bra{0}|)^{\otimes N}]\}$, being $1$ the $2^N\times
2^N$ identity matrix. State $\sigma$ is GHZ-diagonal (see definition in
Ref. \cite{Duer&Cirac}) and all of its negativities are null, then
%Hence, by the D\"ur-Cirac-Tarrach criterion \cite{Duer&Cirac},
$\sigma$ is fully separable \cite{Duer&Cirac}. Consider, for each
qubit $i$, the local POVM $\{A_m^{(i)}\}_{m=1}^2$ with elements
$A_1^{(i)}=\delta(
 \sqrt{p_c^{\text{AD}}(k)}\ket{0}\bra{0}|+\sqrt{1-p_c^{\text{AD}}(k)}\ket{1}\bra{1}|)$, where $\delta$ is such that $A_1^{(i)^{\dagger}} A_1^{(i)}\leq {\bm 1}$,
and $A_2^{(i)^{\dagger}} A_2^{(i)}={\bm 1}-A_1^{(i)^{\dagger}}
A_1^{(i)}$. Applying this POVM to every qubit
of state $\sigma$ yields $\rho_s$ when the measurement outcome is $m=1$
(corresponding to $A_1$) for every qubit.$\square$

%%%%%%%%%%%%%%%%%%%%%%%%%%%%%%%%%%%%%%%%%%%%%%%%%%%%%%%%%%%%%%%%%%%%%%%%%%%%%%%%%%%%%%%%%%%%%%%%%%%%%%%%%%%%%%%%%%%%%%%%%%%%%%%%

%%%%%%%%%%%%%%%%%%%%%%%%%%%%%%%%%%%%%%%%%%%%%%%%%%%%%%%%%%%%%%%%%%%%%%%%%%%%%%%%%%%%%%%%%%%%%%%%%%%%%%%%%%%%%%%%%%%%%%%%%%%%%%%%
%%%%%%%%%%%%%%%%%%%%%%%%%%%%%%%%%%%%%%%%%%%%%%%%%%%%%%%%%%%%%%%%%%%%%%%%%%%%%%%%%%%%%%%%%%%%%%%%%%%%%%%%%%%%%%%%%%%%%%%%%%%%%%%%

\begin{thebibliography}{99}
\bibitem{note} {Since the analized states are
permutationally invariant, $\Lambda_k^{GAD}$ will correspond to the
minimum eigenvalue of the matrix's partial transposition according to
all possible $k:N-k$ partitions. This is also true for the other
channels.}
\end{thebibliography}
\end{document}